\documentclass[%
 aip,
 amsmath,amssymb,
 reprint,%
]{revtex4-1}

\usepackage{graphicx}
\usepackage{dcolumn}
\usepackage{bm}

\usepackage[utf8]{inputenc}
\usepackage[T1]{fontenc}
\usepackage{mathptmx}
\usepackage{xcolor}

\begin{document}

\preprint{AIP/123-QED}

\title[Accepted for publication in Physics of Fluids - Letters, May 2020]{Likelihood of survival of coronavirus in a respiratory droplet deposited on a solid surface}

\author{Rajneesh Bhardwaj}
 \email{rajneesh.bhardwaj@iitb.ac.in.}
 \affiliation{ 
Department of Mechanical Engineering,\\
Indian Institute of Technology Bombay, Mumbai 400076, India
}

\author{Amit Agrawal}%
 \email{amit.agrawal@iitb.ac.in.}
\affiliation{ 
Department of Mechanical Engineering,\\
Indian Institute of Technology Bombay, Mumbai 400076, India
}%


\date{\today}

\begin{abstract}
We predict and analyze the drying time of respiratory droplets from a COVID-19 infected subject, which is a crucial time to infect another subject. The drying of the droplet is predicted by diffusion-limited evaporation \textcolor{black}{model for a sessile droplet placed} on a partially-wetted surface with a pinned contact line. The variation of droplet volume, contact angle, ambient temperature, and humidity are considered. We analyze the chances of the survival of the viruses present in the droplet, based on the lifetime of the droplets in several conditions, and find that the chances of survival of the virus are strongly affected by each of these parameters. The magnitude of shear stress inside the droplet \textcolor{black}{computed using the model} is not large enough to obliterate the virus. We also explore the relationship between the drying time of a droplet and the growth rate of the spread of COVID-19 for five different cities, and find that they are weakly correlated.
\end{abstract}

\maketitle

%

Studies have reported that infectious diseases such as influenza spread through respiratory droplets. The respiratory droplets could transmit the virus from one subject to another through the air. These droplets can be produced by sneezing and coughing. Han et al. \cite{Han2013} measured the size distribution of sneeze droplets exhaled from the mouth. They reported that the geometric mean of the droplet size of 44 sneezes of 20 healthy subjects is around 360 $\mu$m for unimodal distribution and is 74 ${\mu}$m for bimodal distribution. Liu et al. \cite{liu2017evaporation} reported around 20\% \textcolor{black}{longer drying time of saliva droplets as compared to water droplets, deposited on a Teflon-printed slide.} They also predicted and compared these times with a model and considered solute effect (Raoult's effect), due to the presence of salt/electrolytes in saliva. \textcolor{black}{The slower evaporation in the saliva droplet is attributed to the presence of the solute in it \cite{liu2017evaporation}.} Xie et al. \cite{Xie2007} developed a model for estimating the droplet diameter, temperature, and falling distance as a function of time, as droplets are expelled during various respiratory activities. They reported that large droplets expelled horizontally can travel a long distance before hitting the ground. In a recent study, Bourouiba \cite{Bourouiba2020} provided evidence for droplets expelled during sneezing being carried to a much larger distance (of 7-8 m) than previously thought. The warm and moist air surrounding the droplets helps in carrying the droplets to such a large distance.

While the role of virus-laden droplets in spreading infectious diseases is well-known, the drying time of such droplets after falling upon a surface has not been well-studied. {\textcolor{black}{In this context, Buckland and Tyrrell \cite{buckland1962} experimentally studied the loss in infectivity of different viruses upon drying of virus-laden droplets on a glass slide. At  room temperature and 20\% relative humidity, the mean log reduction in titre was reported to be in the range of 0.5-3.7 for the 19 viruses considered by them.}} The need for studying the evaporation dynamics of virus-laden droplets has also been recognized in the recent article by Mittal et al.\cite{mittal2020}. \textcolor{black}{Further, to} reduce the transmission of COVID-19 pandemic, caused by SARS-CoV-2 virus, the use of a face mask has been recommended by WHO \cite{Singhal2020}. The infected droplets could be found on a face mask or a surface inside the room, which necessitates the regular cleaning of the surfaces exposed to droplets. \textcolor{black}{Therefore,} the present study examines the drying times of such droplets which correlates with the time in which the chances of the transmissibility of the virus are high \cite{buckland1962, weber2008}. 

First, we present the different components of the model, used to estimate the drying time and shear stress. We consider aqueous respiratory droplets that are on the order of 1 nL to 10 nL, on a solid surface. The range of the volume is consistent with previous measurements \cite{Han2013}. The corresponding diameters of the droplets in the air are around 125 $\mu$m and 270 $\mu$m and probability density function (PDF) of the normal distribution of the droplet diameter in the air is plotted in Fig. 1. The mean diameter and standard deviation are 188 ${\mu}$m and 42 ${\mu}$m, respectively. Droplets smaller than 100 ${\mu}m$ are not considered in this study because such droplets are expected to remain airborne, while the larger droplets being heavier settle down \cite{Fern2013}. 
\begin{figure}
\includegraphics[width=6cm,keepaspectratio]{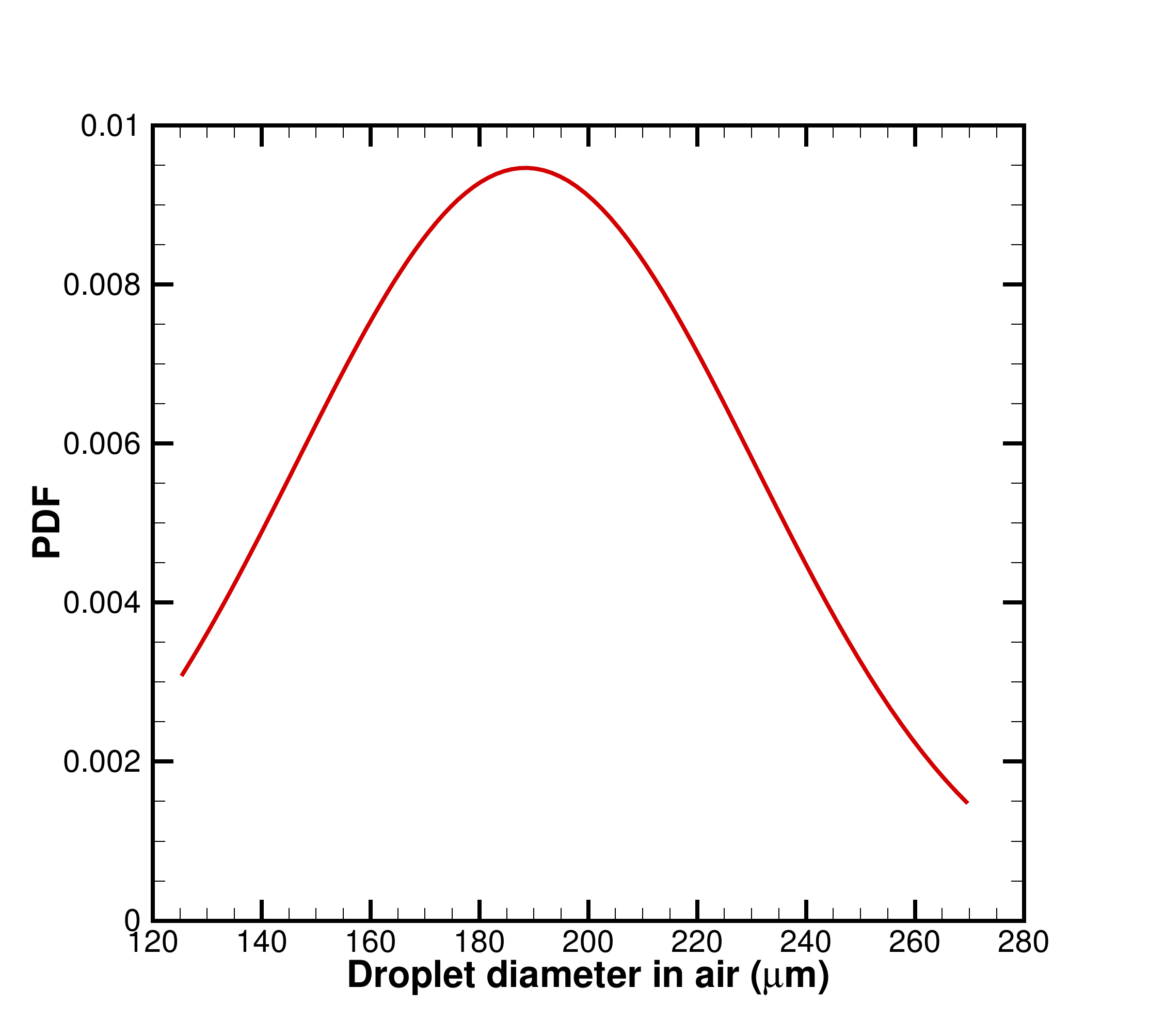}
\caption{\label{fig:PDF}Probability density function (PDF) of the normal distribution of the droplet diameter in air, considered in this letter.}
\end{figure}
The droplet is assumed to be deposited as a spherical cap on the substrate. Since the wetted diameter of the droplet is lesser than capillary length (2.7 mm for water), the droplet maintains a spherical cap shape throughout the evaporation. The volume ($V$) and contact angle ($\theta$) for a spherical cap droplet are   expressed as follows,
\begin{equation}
V={1 \over 6} \pi h(3R^2+h^{2}) \textrm{, } \theta = 2 {\rm tan}^{-1}({h \over R})
\label{eeq13}
\end{equation}
where $h$ and $R$ are droplet height and wetted radius, respectively. We consider diffusion-limited, quasi-steady evaporation of a sessile droplet with a pinned contact line on a partially-wetted surface (Fig.~\ref{fig:sch}). 
\begin{figure}
\includegraphics[width=8cm,keepaspectratio]{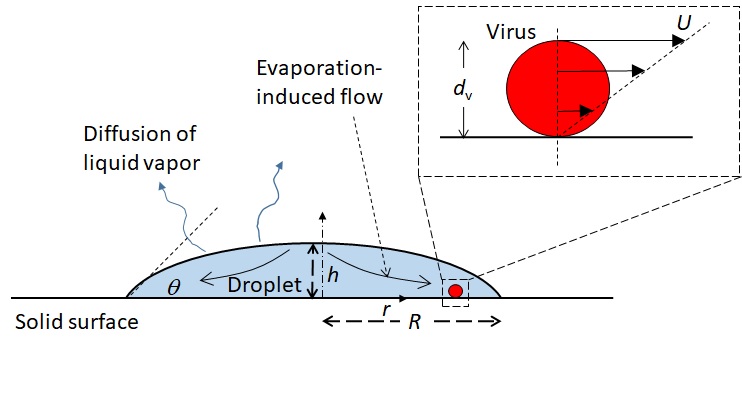}
\caption{\label{fig:sch}Schematic of the problem considered in the present study.}
\end{figure}
The assumption of quasi-steady evaporation is valid for $t_h/t_F < 0.1$, as suggested by Larson \cite{larson2014transport}, where $t_h$ and $t_F$ are heat equilibrium time in the droplet and drying time, respectively. $t_h/t_F$ scales as follows \cite{larson2014transport},  
\begin{equation}
 {t_h \over t_F} \sim 5{D \over \alpha} {h \over R}{c_{sat} \over \rho}
\label{eeq1}
\end{equation}
where $D$, $\alpha$, $h$, $R$, $c_{sat}$ and $\rho$ are diffusion coefficient of liquid vapor in the air, thermal diffusivity of the droplet, droplet height, wetted radius, saturation liquid vapor concentration and droplet density, respectively. In the present work, the maximum value of $t_h/t_F$ is estimated to be around 0.05 at 40$^{\circ}$C, the maximum water droplet temperature considered in the present work and contact angle of 90$^\circ$ ($h/R$ = 1). The values of $D$, $\alpha$ and $\rho$ are taken as $2.5\times10^{-5}$ m$^2$/s, $1.45\times10^{-7}$ m$^2$/s and 997 kg/m$^3$, respectively \cite{lide2004crc}. Therefore, the assumption of quasi-steady evaporation is justified.  

The mass lost rate (kg/s) of an evaporating sessile droplet is expressed as follows \cite{hu2002evaporation},
\begin{equation}
\dot{m} = -{\pi}RD(1-H)c_{sat}(0.27{\theta}^2 + 1.30);
\label{eeq2}
\end{equation}
where, $H$ and $\theta$ are relative humidity and static contact angle, respectively. The saturated concentration (kg/m$^3$) at a given temperature for water vapor is obtained using the following third order polynomial \cite{bhardwaj2009pattern, kumar2018combined}:
\begin{eqnarray}
c_{sat}=9.99\times10^{-7}T^3-6.94\times10^{-5}T^2\nonumber \\
 +3.20\times10^{-3}T-2.87\times10^{-2}
\label{eeq6}
\end{eqnarray}
where $T$ is the temperature in $^\circ$C (20$^\circ$C $\le$ $T$ $<$ 100$^\circ$C). The dependence of diffusion coefficient (m$^2$/s) of water vapor on temperature ($^\circ$C) is given by \cite{bhardwaj2009pattern, kumar2018combined}:
\begin{equation}
D(T)=2.5\times10^{-4}{\rm exp}\left (- {684.15 \over{T+273.15}} \right)
\label{eeq7}
\end{equation}
Assuming a linear rate of change of the volume of the droplet for a sessile droplet pinned on the surface \cite{hu2002evaporation, popov2005evaporative}, the drying time of the droplet is given by, 
\begin{equation}
t_f = {{\rho}V_0 \over \dot{m}}
\label{eeq7}
\end{equation}
where $V_0$ and $\rho$ are the initial volume and density of the droplet. The properties of pure water have been employed in the present calculations to determine the drying time and shear stress. Since the thermo-physical properties of \textcolor{black}{saliva} are not very different from water, the present results provide a good estimate of the evaporation time under different scenarios and shear stress.
Further, we obtain the expression of the maximum shear stress ($\tau$) on 125 nm diameter SARS-CoV-2 virus, suspended in the sessile water droplet and estimate its range for the droplet size considered. The shear stress on the virus would be maximum for a virus adhered to the substrate surface (Fig.~\ref{fig:sch}). Assuming a linear velocity profile across the cross-section of the virus, the expression of $\tau$ given by 
\begin{equation}
\tau = \mu{U \over {d_v}}
\label{eeq7}
\end{equation}
where $\mu$, $U$ and $d_v$ are viscosity of the droplet, flow velocity on the virus apex (Fig.~\ref{fig:sch}) and virus diameter, respectively. The flow inside the droplet is driven by the loss of liquid vapor by diffusion. Previous reports have shown that an evaporating water droplet in ambient does not exhibit Marangoni stresses \cite{hu2005analysis, bhardwaj2009pattern, bhardwaj2010self}, therefore, we estimate $U$ using the evaporative-driven flow. The expression of non-uniform is the evaporative mass flux on the liquid-gas interface, $J$, [kg m$^{-2}$ s$^{-1}$], is given by \cite{hu2002evaporation},  
\begin{eqnarray}
J(r)=&&{Dc_{sat}(1-H)\over{R}}(0.27\theta^2+1.30)\nonumber \\
&&\times(0.6381-0.2239(\theta-\pi/4)^2)(1-(r/R)^2)^{-\lambda(\theta)}
\label{eeq8}
\end{eqnarray}
where $\lambda(\theta)=0.5-\theta/\pi$ and $r$ is radial coordinate (Fig. 1). The above expression exhibits singularity at $r = R$ and the maximum value of $J$ (say $J_{max}$) occurs near the contact line region (say at $r = 0.99R$). The magnitude of the evaporative-driven flow velocity [m s$^{-1}$] is expressed as follows \cite{bhardwaj2010self},
\begin{equation}
U = {J_{max} \over {\rho}}
\label{eeq9}
\end{equation}
The following expression of maximum shear stress ($\tau$) is therefore obtained,
\begin{equation}
\tau = {\mu{U} \over {d_v}} = {\mu{J_{max}} \over {d_v{\rho}}}
\label{eeq10}
\end{equation}
Using Eqs. 8 and 10, the shear stress was estimated on the virus suspended in the droplets of [1-10] nL at $T = 25^{\circ}C$, $\theta = 30^\circ$ and 50\% humidity. \textcolor{black}{To} verify the calculations, we compared the value of $J_{max}$ for a 3.7 nL evaporating water droplet on a glass surface, reported in Ref. \cite{bhardwaj2009pattern}, using finite element simulations. The computed value of $J_{max}$ using eq.~8 is 4.6 $\times$ 10$^{-3}$ kg m$^{-2}$s$^{-1}$, while the value at the contact line in the previous study \cite{bhardwaj2009pattern} is, 5.4 $\times$ 10$^{-3}$ kg m$^{-2}$s$^{-1}$, thereby verifying the present calculations. The computed range of the shear stress is [0.056-0.026] Pa for [1-10] nL droplets. 
%

\begin{figure}
\includegraphics[width=8cm,keepaspectratio]{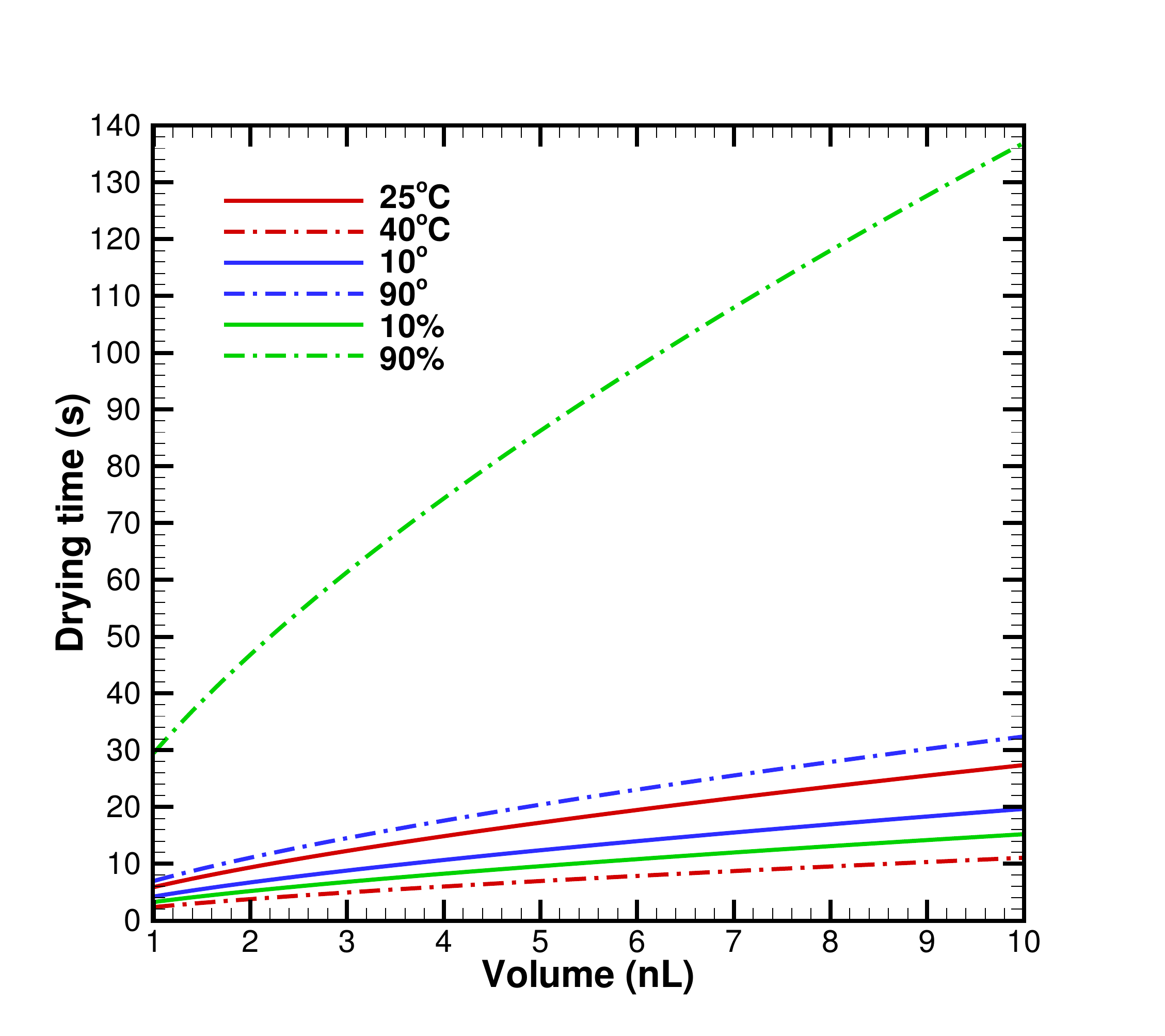}
\caption{\label{fig:slide}Effect of droplet volume on evaporation time, as a function of ambient temperature, surface wettability, and relative humidity.}
\end{figure}

Second, we present the effect of ambient temperature, surface wettability, and relative humidity on the drying time of the droplet.  In this context, we examine the drying time of a deposited droplet in two different ambient temperatures, 25$^{\mathrm{o}}$C and 40$^{\mathrm{o}}$C. The chosen temperatures are representative of temperatures inside a room with air-conditioning and outdoors in summer. Fig.~\ref{fig:slide} shows the variation of evaporation time with droplet volume, at the two different ambient temperatures considered. The contact angle and humidity for these simulations are taken as 30$^{\mathrm{o}}$ and 50\%, respectively. At 25 $^{\mathrm{o}}$C, the evaporation time for small droplets is about 6 s, which increases to 27 s for large size droplets. The evaporation time increases as the square of the droplet radius, or $2/3$ power of volume. An increase in ambient temperature reduces the evaporation time substantially (by about 50\% for 15 $^{\mathrm{o}}$C rise in temperature). Therefore, an increase in the ambient temperature is expected to \textcolor{black}{drastically reduce} the chance of infection through contact with an infected droplet. 
\begin{table}
\caption{\label{table:time}Values of measured contact angle of a water droplet on different surfaces, documented in the literature.}
\begin{ruledtabular}
\begin{tabular}{cccc}
Surface & Contact angle & Study \\
\hline 
Glass & 5-15$^\circ$; 29$^\circ$ & Refs. \cite{roux2004dynamics, kumar2018combined} \\
Wood & 62-74$^\circ$ & Ref. \cite{mantanis1997wetting} \\
Stainless steel & 32$^\circ$ & Ref. \cite{Chandra} \\
Cotton & 41-62$^\circ$ & Ref. \cite{hsieh1996water}\\
Touch screen of smartphone & 74-94$^\circ$ & Ref. \cite{touchscreen}\\
\end{tabular}
\end{ruledtabular}
\end{table}

The effect of the surface on which the droplet can fall \textcolor{black}{onto} is modeled here through an appropriate value of the contact angle. The contact angle of 10$^{\mathrm{o}}$ corresponds to a water droplet on glass, while 90$^{\mathrm{o}}$ corresponds to a water droplet on the touch screen of a smartphone (Table~\ref{table:time}). The results of the simulations corresponding to these two contact angles are plotted in Fig.~\ref{fig:slide}. The ambient temperature and humidity are taken as 25$^{\circ}C$ and 50\%, respectively. Fig.~\ref{fig:slide} shows that the effect of the surface can be quite profound; the evaporation time can increase by 60\% for a more hydrophobic surface. With a decrease in contact angle, the droplet spreads into a thin film, which has a relatively large mass loss rate from the droplet to the ambient. Therefore, for a surface with a smaller contact angle, the evaporation time of the droplet is smaller. The effect of the surface can further manifest by a difference in temperature in different parts of the surface. Such inhomogeneity in surface temperature can be brought about by the difference in the surface material (leading to the difference in the emissivity) or differential cooling (for example, due to the corner effect). Even a slight difference in the surface temperature can further aggravate the surface effect by influencing the evaporation time. 

The SARS-CoV-2 virus has a lipid envelop, and in general, the survival tendency of such viruses, \textcolor{black}{when suspended in air,} is larger at a lower relative humidity of 20–30\% \cite{Tang2009}, \textcolor{black}{as compared to several other viruses which do not have a protective lipid layer. Here, we examine the effect of the relative humidity on the survival of the virus inside a droplet, deposited on a surface.} Fig.~\ref{fig:slide} shows that the relative humidity has a strong effect on the evaporation time. \textcolor{black}{The contact angle and ambient temperature for these calculations are taken as 30$^{\mathrm{o}}$ and 25$^{\mathrm{o}}$C, respectively.} The evaporation time of a droplet increases almost 7-fold with an increase in humidity from 10\% to 90\%. Further, the evaporation time becomes greater than 2 min for large droplets at high humidity. With the increase in humidity in coastal areas in summer and later in other parts of Asia in July-September with advent of Monsoon, this may become an issue, as there will be sufficient time for the virus to spread from the droplet to new hosts, upon contact with the infected droplet. \textcolor{black}{Therefore, a larger humidity increases the survival of the virus when it is inside the droplet, however, it decreases its chances of the survival if the virus is airborne.}

Finally, we discuss the relevance of the present results in the context of COVID-19 pandemic. The evaporation time of a droplet is a critical parameter as it determines the duration over which spread of infection from the droplet to another person coming in contact with the droplet, is possible.  The virus needs a medium to stay alive \cite{buckland1962}; therefore, once the droplet has evaporated, the virus is not expected to survive.  The evaporation time can, therefore, be taken as an indicator of the survival time of the virus. In general, it is regarded that a temperature of 60 $^{\mathrm{o}}$C maintained for more than 60 min inactivates most viruses \cite{Tang2009}; however, contrary reports about the effect of temperature on the survivability of SARS-CoV-2 virus has been reported. \cite{Wang2020, Yao2020} Our results indicate that the survival time of the virus depends on the surface on which the droplet has fallen, along with the temperature and humidity of the ambient air. \begin{figure}
\includegraphics[width=8cm,keepaspectratio]{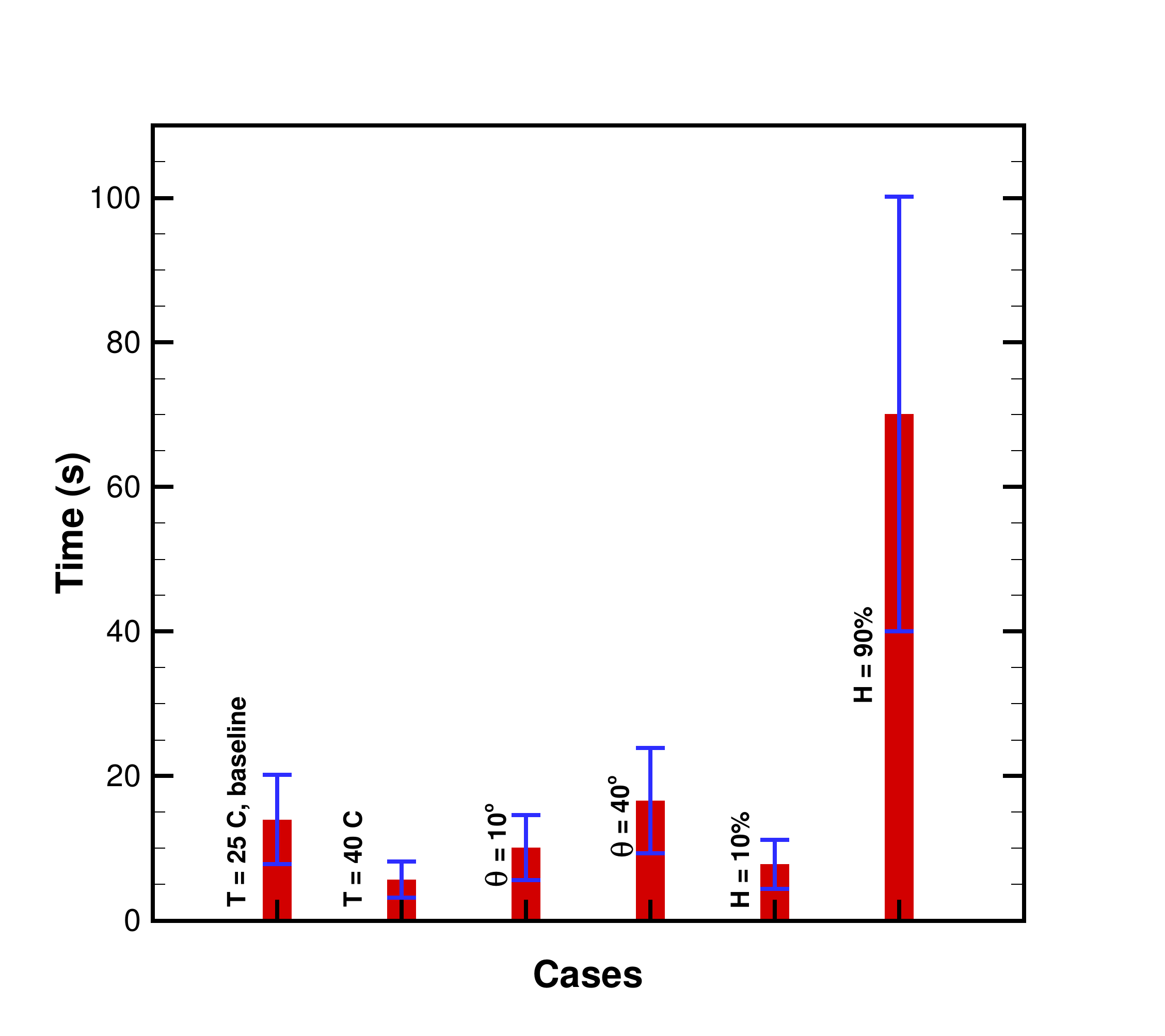}
\caption{\label{SD}Mean and standard deviation of the probability density function of computed drying time normal distribution. The drying time was calculated for the droplet volume distribution, plotted in Fig.~\ref{fig:PDF}. The mean and standard deviation are shown by a vertical red bar and error bar, respectively, for different cases considered in the study.}
\end{figure}
The present results are expected to be of relevance in two different scenarios: When droplets are generated by an infected person by coughing or sneezing (in the absence of a protective mask), or when fine droplets are sprayed on a surface for cleaning/disinfecting the surface. A wide range of droplet sizes is expected to be produced in these cases. The mutual interaction of the droplets, such that they interfere in the evaporation dynamics, is however expected to be weak because of the large distance between the droplets as compared to their diameter. 

The virus inside a droplet is subjected to shear stresses, due to the generation of evaporation-induced flow inside the droplet. The magnitude of this shear stress has however been estimated to be small and the virus is unlikely to be disrupted by this shear stress inside the droplet.

\textcolor{black}{To} determine the likelihood of the droplet and the virus on the surface, we find the mean and standard deviation of the probability density function (PDF) of the normal distribution of the droplet drying times for different cases of ambient temperature, contact angle and relative humidity.
The values of the mean and standard deviation are plotted using bar and error bar, respectively, in Fig.~\ref{SD}. The likelihood lifetime is in the range for [5-20] s for $H \le 50\%$ while it is a range of [40-100] s for $H = 50\%$. This result shows that the drying time is likely to be larger by around five times in case of large relative humidity values, thereby, increasing the chances of the survival of the virus. 

\begin{figure}
\includegraphics[width=8cm,keepaspectratio]{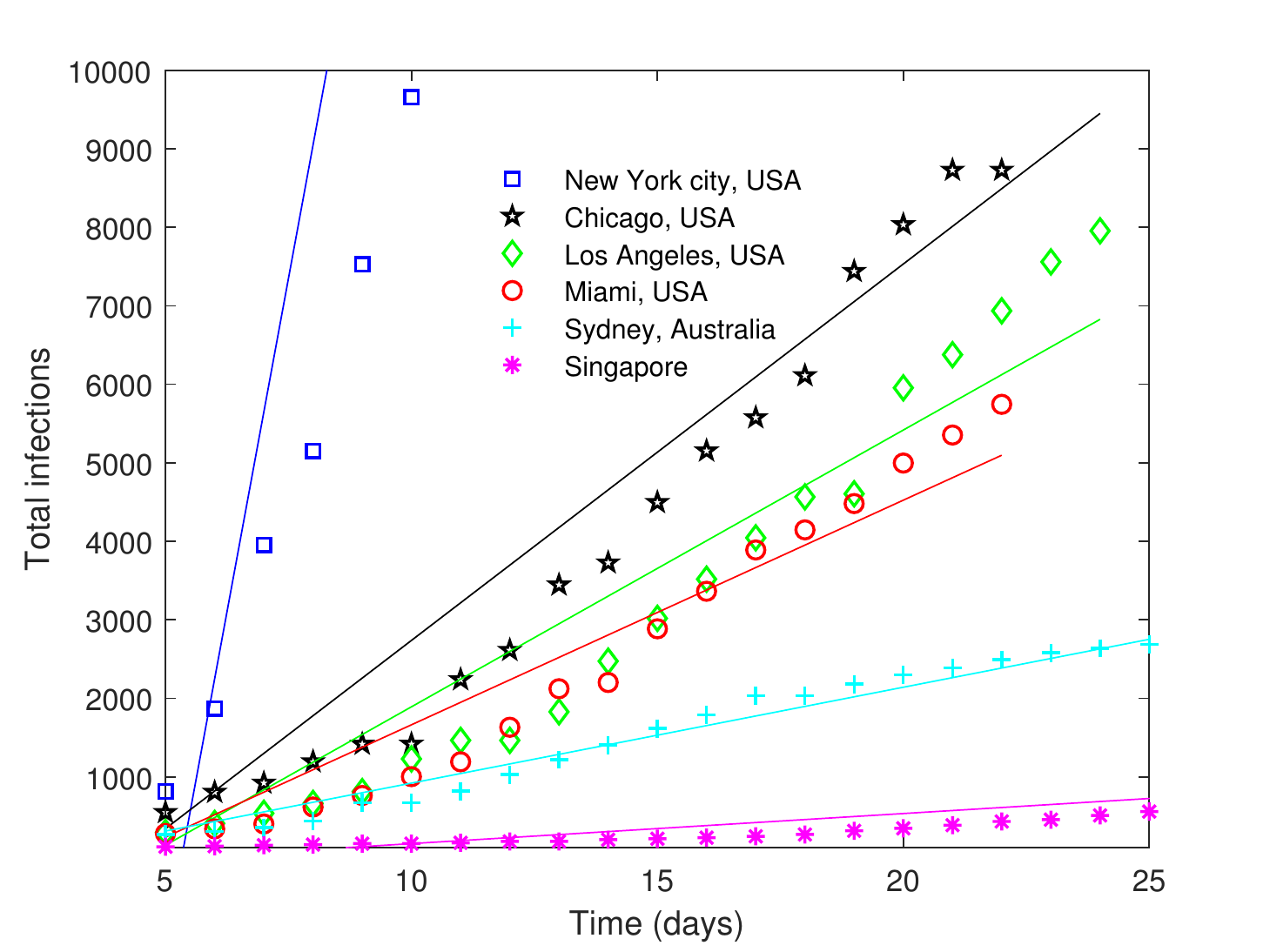}
\caption{\label{cities}Comparison among evolution of the total infections of different cities/regions. Day 0 is defined as the day on which the total number of infections is 100 or larger. The slope of the linear fit obtained using least-squares method is considered as the growth rate of the infection \textcolor{black}{(number of infections per day)}.}
\end{figure}

\begin{table}
\caption{\label{table2} Approximate range of outdoor ambient temperature and relative humidity during the duration of pandemic (1 March 2020 to 10 April 2020) in different cities/regions. The data is compiled from Ref. \cite{w3}.}
\begin{ruledtabular}
\begin{tabular}{ccc}
City/Region & Ambient temperature & Relative humidity \\
\hline 
New York City  & 6-10$^{\circ}$C    &  50-60\%  \\
Chicago        & 4-8$^{\circ}$C     &  60-70\%  \\
Los Angeles    & 14-18$^{\circ}$C   &  45-55\%  \\
Miami          & 20-24$^{\circ}$C   &  65-75\%  \\
Sydney         & 21-25$^{\circ}$C   &  55-65\%  \\
Singapore      & 28-32$^{\circ}$C   &  70-80\%  
\end{tabular}
\end{ruledtabular}
\end{table}
\begin{figure}
\includegraphics[width=8cm,keepaspectratio]{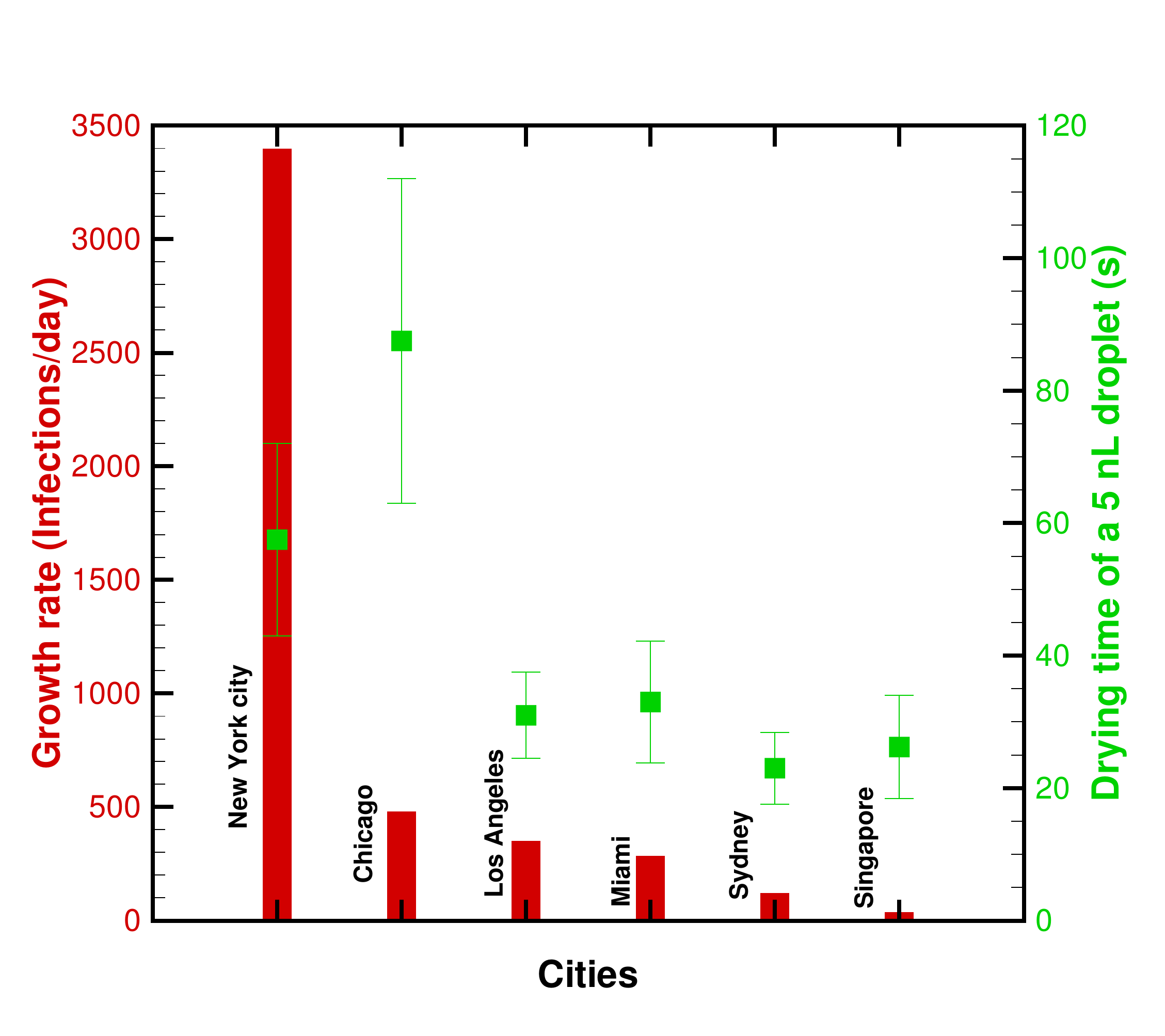}
\caption{\label{growth}Comparison of the growth rate of the infection of different cities/regions (bars) with respective drying times (squares) of a 5 nL droplet. The error bar represents the variability in outdoor weather.}
\end{figure}
Further, we examine the connection between the drying time of a droplet to the growth of the infection. A similar approach was tested for suspended droplets in air in Ref. \cite{basu} recently. We hypothesize that since the drying time of a respiratory droplet on a surface is linked to the survival of the droplet, it is correlated with the growth of the pandemic. Since the drying time is a function of weather, we compare the growth of infection with the drying time for different cities. \textcolor{black}{The cities were selected based on cold/warm and dry/humid weather.} The growth of the total number of infections is plotted for cities with different weather during the pandemic in Fig.~\ref{cities}. The data of the infections were obtained from public repositories \cite{w1,w2}. The data were fitted with linear curves using the least-squares method and the slope of the fits represents the growth rate \textcolor{black}{(number of infections per day)} of the respective city. The growth rate of New York City and Singapore is the largest and lowest, respectively. 

For different cities, we compute the drying time of a droplet of 5 nL volume, which is the mean volume obtained using PDF of the distribution (Fig.~\ref{fig:PDF}). The ambient temperature and relative humidity are taken as mean of the respective ranges, listed in Table~\ref{table2}. \textcolor{black}{As discussed earlier, the drying time increases with increase in humidity, however, it decreases with increase in ambient temperature. Thus, the combined effect of  humidity and temperature decides the final drying time. This can be illustrated by comparing the drying time of Singapore and New York city, plotted in Fig.~\ref{growth}. The time is shorter for the former as compared to the latter, despite with a large humidity for the former (70-80\%) as compared to the latter (50-60\%). } 

\textcolor{black}{Lastly, Fig.~\ref{growth} compares} the growth rate and drying time of the different cities using vertical bars and symbols, respectively. The growth rate appears to be weakly correlated with the drying time i.e. a larger (lower) growth rate corresponds to larger (lower) drying time. Qualitatively, this data verifies that when a droplet evaporates slowly, the chance for survival of the virus is enhanced and the growth rate is augmented.

We recognize that there are limitations of the model presented here, which can be improved in subsequent studies. In particular, the ambient air has been assumed to be stationary; the evaporation time is expected to reduce in presence of convective currents. Therefore, the value of the predicted evaporation times is on the conservative side and the actual evaporation time will be smaller than that obtained here. The effect of the solute present (i.e., Raoult's law) in saliva/mucus has not been modeled and the contact angle and drying of these biological fluids could be slightly different from that of pure water on a solid surface. However, the impact of these latter effects on the drying time is expected to be small. \textcolor{black}{Further, the model does not consider the interaction of the droplets. It is likely that the respiratory droplets, expelled from mouth and/or nose, deposit adjacent to each other on a surface and could interact while evaporating \cite{Sen2020}. They may interact while falling \cite{Giulia} and a falling droplet may coalesce on an already deposited droplet on a surface \cite{kumar2020}. In addition, receding of the contact line may influence the drying time \cite{Stauber}, which is not considered in the present work.} 

In closure, we have examined the likelihood of survival of the SARS-CoV-2 virus suspended in respiratory droplets originated from a COVID-19 infected subject. The droplet considered to be evaporating in a quiet ambient on different surfaces. The droplets volume range is considered as [1, 10] nL. The datasets of drying time presented here for different ambient conditions and surfaces will be helpful in future studies. The likelihood of the survival of the virus increases roughly by 5 times in a humid ambient as \textcolor{black}{compared} to a dry ambient. The growth rate of the COVID-19 was found to be weakly correlated with the outdoor weather. \textcolor{black}{While the present Letter discusses the results in the context of COVID-19, the present model is also valid for respiratory droplets of other transmissible diseases, such as Influenza A.}

\begin{acknowledgments}
R.B. gratefully acknowledges financial support by a grant (EMR/2016/006326) from Science and Engineering Research Board (SERB), Department of Science and Technology (DST), New Delhi, India.
\end{acknowledgments}
\vspace{5mm}
\noindent\textbf{Data Availability Statement}\\
The data that support the findings of this study are available from the corresponding author(s) upon reasonable request.

\nocite{*}
\bibliography{aipsamp}

\end{document}